\documentclass[12pt,thmsa]{article}
\usepackage{sw20aip}

\input{tcilatex}
\begin{document}

\title{MAGNETIC RESPONSE IN ANYON FLUID AT HIGH TEMPERATURE\thanks{%
This work has been supported in part by NSF grant PHY-9722059}}
\author{E. J. Ferrer and V. de la Incera \\
Dept. of Phys., SUNY, Fredonia, NY 14063, USA\\
SUNY-FRE-98-06 \and {\small Talk presented at SILAFAE'98. April 8-11, 1998,
San Juan, Puerto Rico }}
\maketitle

\begin{abstract}
The magnetic response of the charged anyon fluid at temperatures lower and
larger than the fermion enery gap $\omega _{c}$ is investigated in the
self-consistent field approximation. We prove that the anyon system with
boundaries exhibits a total Meissner effect at temperatures smaller than the
fermion energy gap ($T\ll \omega _{c}$). The London penetration length at $T$
$\sim 200K$ is of the order $\lambda \sim 10^{-5}cm$. At $T\gg \omega _{c}$
a new phase, characterized by an inhomogeneous magnetic penetration, is
found. We conclude that the energy gap, $\omega _{c},$ defines a scale that
separates two phases: a superconducting phase at $T\ll \omega _{c}$, and a
non-superconducting one at $T\gg \omega _{c}$.
\end{abstract}

Anyons \cite{Anyons}$^{,}$ \cite{Wilczek} are particles with fractional
statistics in (2+1)-dimensions. The anyon description within the
Chern-Simons (CS) gauge theory is equivalent to attaching flux tubes to the
charged fermions. The Aharonov-Bohm phases resulting from the adiabatic
transport of two anyons is the source of the fractional exchange statistics%
\cite{Wilczek}$.$

It has been argued that strongly correlated electron systems in two
dimensions can be described by an effective field theory of anyons \cite{5}$%
. $ Anyons can be also obtained as solitons with fractional spin in electron
systems. Excitations with fractional spin in two dimensional systems
necessarily obey fractional statistics \cite{5aa}$.$

As it is known, anyon superconductivity has an origin different from the
Nambu-Goldstone-Higgs like mechanism. The genesis of the anyon
superconductivity is given by the spontaneously violation of commutativity
of translations in the free anyon system\cite{Chen}$.$ This new mechanism
might find wide applications in new physical studies.

Anyon superconductivity at $T=0$ is a well establish result \cite{Arova}$%
^{,} $ \cite{Chen}$.$ However, at $T\neq 0$ several authors \cite{9} have
advocated that the superconducting phase evaporates at any finite
temperature. The reasons is that at $T\neq 0$ there exists a long-range
electromagnetic mode inside the infinite bulk \cite{8}$.$ This long range
mode is the consequence of the existence of a pole $\sim \left( \frac{1}{%
\mathbf{k}^{2}}\right) $ in the fermion polarization operator component $\Pi
_{00}$ at finite temperature.

In previous works \cite{Our} we found that, contrary to some authors'
belief, the superconducting behavior, manifested through the Meissner effect
in the charged anyon fluid at $T=0$, does not disappear as soon as the
system is heated. In Ref. [10] we showed that the presence of boundaries
affects the dynamics of the two-dimensional system in such a way that the
long-range mode, that accounts for a homogeneous field penetration \cite{8}$%
, $ cannot propagate in the bulk. According to these results, the anyon
system with boundaries exhibits a total Meissner effect at temperatures
smaller than the fermion energy gap ($T\ll \omega _{c}$).

It is natural to expect that at temperatures larger than the energy gap this
superconducting behavior should not exist. At those temperatures the
electron thermal fluctuations should make accessible the free states
existing beyond the energy gap. These heuristic arguments were corroborated
in Ref. [11]$.$ There, we proved that at $T\gg \omega _{c}$ the charged
anyon fluid does not exhibit a Meissner effect.

We can conclude that the energy gap $\omega _{c}$ defines a scale that
separates two phases in the charged anyon fluid: a superconducting phase at $%
T\ll \omega _{c}$, and a non-superconducting one at $T\gg \omega _{c}$.

We must emphasize that the scenario we have found for the anyon
superconductivity at finite temperature is in agreement with the rationale
of anyon superconductivity given by Wilczek\cite{Wilczek}$.$ Wilczek has
pointed out that the London arguments, which start from the role of the
energy gap as an essential fact in the theory of superconductivity, seem to
provide the base for anyon superconductivity. In the charged anyon fluid,
there is no a charge-violating local order parameter so familiar in the
theories with spontaneously broken symmetry. In this system, instead, it is
the background CS magnetic field $\overline{b}$ what determines the energy
gap ($\omega _{c}=\overline{b}/m$) and plays the role of the order parameter
in the anyon gas \cite{Chen}$.$

In what follows we present the results for the linear magnetic response of
the charged anyon fluid to an applied constant and uniform magnetic field,
at temperatures lower and larger than the energy gap.

The linear response of the medium can be found under the assumption that the
quantum fluctuations of the gauge fields about the ground-state are small.
In this case the one-loop fermion contribution to the effective action,
obtained after integrating out the fermion fields, can be evaluated up to
second order in the gauge fields. The effective action in terms of the
quantum fluctuation of the gauge fields within the linear approximation \cite
{8}$^{,}$\cite{11} takes the form

\begin{equation}
\Gamma _{eff}\,\left( A_{\nu },a_{\nu }\right) =\int dx\left( -\frac{1}{4}%
F_{\mu \nu }^{2}-\frac{N}{4\pi }\varepsilon ^{\mu \nu \rho }a_{\mu }\partial
_{\nu }a_{\rho }\right) +\Gamma ^{\left( 2\right) }  \tag{1}
\end{equation}
$\Gamma ^{\left( 2\right) }$ is the one-loop fermion contribution to the
effective action in the linear approximation

\begin{equation}
\Gamma ^{\left( 2\right) }=\int dxdy\left[ a_{\mu }\left( x\right) +eA_{\mu
}\left( x\right) \right] \Pi ^{\mu \nu }\left( x,y\right) \left[ a_{\nu
}\left( y\right) +eA_{\nu }\left( y\right) \right] .  \tag{2}
\end{equation}

In (2) $\Pi _{\mu \nu }$ represents the fermion one-loop polarization
operator in the presence of the CS background magnetic field $\overline{b}$.

Taking into account that we will investigate the magnetic response of the
charged anyon fluid to a uniform and constant applied magnetic field, we
need the $\Pi _{\mu \nu }$ leading behavior for static $\left(
k_{0}=0\right) $ and slowly $\left( \mathbf{k}\sim 0\right) $ varying
configurations. In this limit, and using the frame on which $k^{i}=\left(
k,0\right) $, $i=1,2$, the polarization operator takes the form

\begin{equation}
\Pi ^{\mu \nu }=\left( 
\begin{array}{ccc}
-\left( \mathit{\Pi }_{\mathit{0}}+\mathit{\Pi }_{\mathit{0}}\,^{\prime
}\,k^{2}\right) & 0 & -i\mathit{\Pi }_{\mathit{1}}k \\ 
0 & 0 & 0 \\ 
i\mathit{\Pi }_{\mathit{1}}k & 0 & \mathit{\Pi }_{\,\mathit{2}}k^{2}
\end{array}
\right)  \tag{3}
\end{equation}

The leading contributions of the one-loop polarization operator coefficients 
$\mathit{\Pi }_{\mathit{0}}$, $\mathit{\Pi }_{\mathit{0}}\,^{\prime }$, $%
\mathit{\Pi }_{\mathit{1}}$ and $\mathit{\Pi }_{\,\mathit{2}}$ in the static
limit ($k_{0}=0,$ $\mathbf{k}\sim 0$) at low temperatures $\left( T\ll
\omega _{c}\right) $ are

\[
\mathit{\Pi }_{\mathit{0}}=\frac{2\beta \overline{b}}{\pi }e^{-\beta 
\overline{b}/2m},\qquad \mathit{\Pi }_{\mathit{0}}\,^{\prime }=\frac{mN}{%
2\pi \overline{b}}\mathit{\Lambda },\qquad \mathit{\Pi }_{\mathit{1}}=\frac{N%
}{2\pi }\mathit{\Lambda },\quad \mathit{\Pi }_{\,\mathit{2}}=\frac{N^{2}}{%
4\pi m}\mathit{\Lambda }^{\prime }, 
\]

\begin{equation}
\mathit{\Lambda }=\left[ 1-\frac{2\beta \overline{b}}{m}e^{-\beta \overline{b%
}/2m}\right] ,\quad \mathit{\Lambda }^{\prime }=\left[ \mathit{\Lambda }-%
\frac{2\beta \overline{b}}{mN^{2}}e^{-\beta \overline{b}/2m}\right]  \tag{4}
\end{equation}
and at high temperatures $\left( T\gg \omega _{c}\right) $ are

\[
\mathit{\Pi }_{\mathit{0}}=\frac{m}{2\pi }\left[ \tanh \frac{\beta \mu }{2}%
+1\right] ,\quad \mathit{\Pi }_{\mathit{0}}\,^{\prime }=-\frac{\beta }{48\pi 
}\func{sech}\!^{2}\!\,\left( \frac{\beta \mu }{2}\right) ,\quad \mathit{\Pi }%
_{\mathit{1}}=\frac{\overline{b}}{m}\mathit{\Pi }_{\mathit{0}}\,^{\prime
},\quad 
\]

\begin{equation}
\mathit{\Pi }_{\,\mathit{2}}=\frac{1}{12m^{2}}\mathit{\Pi }_{\mathit{0}} 
\tag{5}
\end{equation}
In these expressions $\mu $ is the chemical potential and $m=2m_{e}$ ($m_{e}$
is the electron mass). These results are in agreement with those found in
Ref.[8]$.$

The extremum equations obtained from the effective action (1) for the
Maxwell and CS fields are

\[
\mathbf{\nabla }\cdot \mathbf{E}=eJ_{0},\qquad -\partial
_{0}E^{k}+\varepsilon ^{kl}\partial _{l}B=eJ^{k} 
\]

\begin{equation}
\frac{eN}{2\pi }b=\mathbf{\nabla }\cdot \mathbf{E,\qquad }\frac{eN}{2\pi }%
f_{0k}=\varepsilon _{kl}\partial _{0}E^{l}+\partial _{k}B  \tag{6}
\end{equation}
$f_{\mu \nu }$ is the CS gauge field strength tensor, defined as $f_{\mu \nu
}=\partial _{\mu }a_{\nu }-\partial _{\nu }a_{\mu }$, and $J_{ind}^{\mu }$
is the current density induced by the many-particle system.

\begin{equation}
J_{ind}^{0}\left( x\right) =\mathit{\Pi }_{\mathit{0}}\left[ a_{0}\left(
x\right) +eA_{0}\left( x\right) \right] +\mathit{\Pi }_{\mathit{0}%
}\,^{\prime }\partial _{x}\left( \mathcal{E}+eE\right) +\mathit{\Pi }_{%
\mathit{1}}\left( b+eB\right)  \tag{7}
\end{equation}

\begin{equation}
J_{ind}^{1}\left( x\right) =0,\qquad J_{ind}^{2}\left( x\right) =\mathit{\Pi 
}_{\mathit{1}}\left( \mathcal{E}+eE\right) +\mathit{\Pi }_{\,\mathit{2}%
}\partial _{x}\left( b+eB\right)  \tag{8}
\end{equation}
In the above expressions we used the following notation: $\mathcal{E}=f_{01}$%
, $E=F_{01}$, $b=f_{12}$ and $B=F_{12}$. We confine our analysis to gauge
field configurations which are static and uniform in the $y$-direction.
Within this restriction we are taking a gauge in which $A_{1}=a_{1}=0$.

The magnetic field solution obtained from eqs. (6)-(8) is

\begin{equation}
B\left( x\right) =-\gamma _{1}\left( C_{1}e^{-x\xi _{1}}-C_{2}e^{x\xi
_{1}}\right) -\gamma _{2}\left( C_{3}e^{-x\xi _{2}}-C_{4}e^{x\xi
_{2}}\right) +C_{5}  \tag{9}
\end{equation}
where $\gamma _{1}=\left( \xi _{1}^{2}\kappa +\eta \right) /\xi _{1}$, $%
\gamma _{2}=\left( \xi _{2}^{2}\kappa +\eta \right) /\xi _{2}$, $\kappa =%
\frac{2\pi }{N\delta }\mathit{\Pi }_{\,\mathit{2}}$, $\eta =-\frac{e^{2}}{%
\delta }\mathit{\Pi }_{\mathit{1}}$. As can be seen from the magnetic field
solution (9), the real character of the inverse length scales $\xi _{1}$ and 
$\xi _{2}$ is crucial for the realization of the Meissner effect.

At temperatures much lower than the energy gap $\left( T\ll \omega
_{c}\right) $ the inverse length scales are given by the following real
functions 
\begin{equation}
\xi _{1}\simeq \sqrt{\frac{e^{2}\mathit{\Pi }_{\mathit{1}}}{\pi \mathit{\Pi }%
_{\,\mathit{2}}}}=e\sqrt{\frac{m}{\pi }}\left[ 1+\frac{\pi n_{e}}{2m}\beta
\exp -\left( \frac{\pi n_{e}\beta }{2m}\right) \right]  \tag{10}
\end{equation}
\begin{equation}
\xi _{2}\simeq \sqrt{\frac{e^{2}\mathit{\Pi }_{\mathit{1}}}{\pi \mathit{\Pi }%
_{\mathit{0}}\,^{\prime }}+\frac{\mathit{\Pi }_{\mathit{0}}}{\mathit{\Pi }_{%
\mathit{0}}\,^{\prime }}}=e\sqrt{\frac{n_{e}}{m}}\left[ 1+\frac{\pi ^{2}n_{e}%
}{e^{2}}\beta \exp -\left( \frac{\pi n_{e}\beta }{2m}\right) \right] 
\tag{11}
\end{equation}

While at temperatures much larger than the energy gap $\left( T\gg \omega
_{c}\right) $ the inverse length scales are given by

\begin{equation}
\xi _{1}\simeq e\sqrt{\mathit{\Pi }_{\mathit{0}}}=e\sqrt{m/2\pi }\left(
\tanh \frac{\beta \mu }{2}+1\right) ^{\frac{1}{2}}  \tag{12}
\end{equation}

\begin{equation}
\xi _{2}\simeq \frac{1}{\pi }\left( \mathit{\Pi }_{\,\mathit{2}}\mathit{\Pi }%
_{\mathit{0}}\,^{\prime }\right) ^{-1/2}=24i\sqrt{2m/\beta }\cosh \frac{%
\beta \mu }{2}\left( \tanh \frac{\beta \mu }{2}+1\right) ^{-\frac{1}{2}} 
\tag{13}
\end{equation}

The imaginary value of the inverse length $\xi _{2}$ at $\left( T\gg \omega
_{c}\right) $ is due to the fact that at those temperatures, $\mathit{\Pi }%
_{\,\mathit{2}}>0$ and $\mathit{\Pi }_{\mathit{0}}\,^{\prime }<0$ (see eq.
(5)). An imaginary $\xi _{2}$ implies that the term $\gamma _{2}\left(
C_{3}e^{-x\xi _{2}}-C_{4}e^{x\xi _{2}}\right) $, in the magnetic field
solution (9), does not have a damping behavior, but an oscillating one.

On the other hand, the presence of the constant coefficient $C_{5}$ in the
magnetic field solution (9) means that there exists a magnetic long-range
mode. Nevertheless, to completely determine the characteristics of the
magnetic response in this case, it is needed to find the values of the $%
C^{\prime }s$ unknown coefficients which are in agreement with the problem
boundary conditions and the minimization of the system free-energy density.
Considering that the anyon fluid is confined to a half plane $-\infty
<y<\infty $ with boundary at $x=0$. The boundary conditions for the magnetic
field are $B\left( x=0\right) =\overline{B}$ ($\overline{B}$ constant), and $%
B\left( x\rightarrow \infty \right) $ finite. Because no external electric
field is applied, the boundary conditions for this field are, $E\left(
x=0\right) =0$, $E\left( x\rightarrow \infty \right) $ finite. The leading
contribution of the stable magnetic configurations which satisfy the problem
boundary conditions are given at $T\ll \omega _{c}$ by\cite{Our}

\begin{equation}
B\left( x\right) =\overline{B}e^{-\xi _{2}x}  \tag{14}
\end{equation}
while at $T\gg \omega _{c}$ it is\cite{New}

\begin{equation}
B\left( x\right) =\overline{B}\cos \left( \left| \xi _{2}\right| x\right) 
\tag{15}
\end{equation}

From (14) and (11) we have that at temperatures lower than the energy gap ($%
T\ll \omega _{c}$) a constant and uniform applied magnetic field cannot
penetrate the anyon fluid (i.e. the Meissner effect takes place in that
superconducting phase) since it exponentially decays with a London
penetration length $\lambda \sim 10^{-5}cm$ at $T$ $\sim 200K$. On the other
hand, from (15) and (13) we have that at $T\gg \omega _{c}$ there is not
Meissner effect, but an inhomogeneous magnetic penetration. Hence, we can
conclude that the energy gap $\omega _{c}$ defines a scale that separates
two phases in the charged anyon fluid: a superconducting phase at $T\ll
\omega _{c}$, and a non-superconducting one at $T\gg \omega _{c}$.

\end{document}